\documentclass[twocolumn,superscriptaddress, showpacs] {revtex4}
\usepackage{graphicx}
\usepackage{amsmath}
\usepackage{dsfont}
\usepackage{amssymb}
\usepackage{color}
\begin{document}

\title[Short Title]{Stationary three-dimensional entanglement  via dissipative Rydberg pumping}
\author{Xiao-Qiang Shao\footnote{Corresponding author: xqshao@yahoo.com}}
\affiliation{School of Physics, Northeast Normal University,
Changchun 130024, People's Republic of China}
\affiliation{Centre for Quantum Technologies, National University of Singapore, 3 Science Drive 2, Singapore 117543}
\author{Jia-Bin You}
\affiliation{Centre for Quantum Technologies, National University of Singapore, 3 Science Drive 2, Singapore 117543}
\author{Tai-Yu Zheng}
\affiliation{School of Physics, Northeast Normal University,
Changchun 130024, People's Republic of China}
\author{C. H. Oh}
\affiliation{Centre for Quantum Technologies, National University of Singapore, 3 Science Drive 2, Singapore 117543}
\author{Shou Zhang}
 \affiliation{Department of Physics, College of Science,
Yanbian University, Yanji, Jilin 133002, People's Republic of China}
\begin{abstract}
We extend the recent result of  bipartite Bell singlet [Carr and Saffman, Phys. Rev. Lett. {\bf111}, (2013)]  to a stationary three-dimensional entanglement between two-individual neutral Rydberg atoms. This proposal makes full use of the coherent dynamics provided by Rydberg mediated interaction and the dissipative factor originating from the spontaneous emission of Rydberg state. The numerical simulation of the master equation reveals that both the target state negativity ${\cal N}(\hat\rho_{\infty})$ and fidelity ${\cal F(\hat\rho_{\infty})}$ can exceed 99.90\%. Furthermore, a steady three-atom singlet state $|S_3\rangle$ is also achievable based on the same mechanism.
\end{abstract}
 \maketitle \maketitle
\section{introduction}
Dissipation has long been regarded as one major obstacle to developing any quantum technology in experiment. Indeed, the quantum information, stored in certain quantum system, will be lost due to the inevitable interaction between quantum system and its surroundings. This fact has degraded many quantum entanglements and quantum logic operations relying on unitary dynamics, and impels
people to explore efficient technologies to confront decoherence, such as quantum error correction \cite{chi,kosut,reed}, decoherence-free subspace \cite{lidar,beige,kempe},  quantum controls \cite{gu,car,xue}, structured environments \cite{aa}, classical environmental driving \cite{bb}, suitable non-dissipative channels \cite{cc} and so on. Although there are some proposals on the production of noise-assisted entanglement \cite{yi}, the entanglement is too small to be utilized.  In 2009,
Verstraete {\it et al.}  proposed a novel scheme for quantum computation and quantum-state engineering driven by dissipation \cite{verstraete}, where the coupling to the environment drived the system to a target steady state. This work has shed new light on quantum information processing since it illustrates that the dissipation can be utilized as a resource to prepare entanglement and implement universal
quantum computing. Since then, the dissipation-based protocols have sprung up in
various quantum information tasks \cite{vacanti,kastoryano,voll,busch,shen,cano,pengbo,shao,torre,rao1,carr,lin,leg,lee,shao1,otter}. For instance, Kastoryano {\it et al}. prepared a maximally bipartite-entanglement state in a leaking optical cavity \cite{kastoryano}. Dalla Torre {\it et al}. realized the spin squeezing using quantum-bath engineering in a dissipative atom-cavity system \cite{torre}. Leghtas {\it et al}. also prepared and protected a maximally entangled state of a pair of superconducting qubits in a low-$Q$ cavity \cite{leg}. Compared with unitary-based scenarios, the main advantage for adopting dissipation is that a stationary desired state can be achievable irrespective to the initial one.

Among many physical carriers of qubit, neutral atoms are good candidates for quantum information processing because they possess stable hyperfine ground states fitting for encoding and any pairs of long-range atoms are coupled to each other via Rydberg mediated interaction \cite{saff,jaks,urban}. As this interaction is
strong enough to shift the atomic energy levels of highly excited Rydberg
states, the Rydberg blockade occurs and the population of bi-excitation state is restrained. Until now, the Rydberg blockade has been successfully applied in
quantum computation \cite{dm,ms,isen,wu,rao}, preparation of entangled state \cite{saff1,zhang,wilk,sn}, quantum algorithms \cite{chen}, quantum simulators \cite{weimer}, and quantum repeaters \cite{han}. In
this paper, we extend the recent work of Carr and Saffman \cite{carr} to preparation of a three-dimensional entanglement of Rydberg atoms. This kind of entangled state can enhance the violations of local realism and the security of quantum cryptography. The proposal makes full use of the unitary dynamics provided by Rydberg interaction and the dissipative factor originating from the spontaneous emission of Rydberg state. Our result shows that  both the target state negativity ${\cal N}(\hat\rho_{\infty})$ and fidelity ${\cal F}(\hat\rho_{\infty})$ can exceed 99.90\%, which outperforms the schemes according to unitary dynamics \cite{ye,lu,li,libo}.

The structure of the paper is organized  as follows. We first briefly review the advantage of the multi-dimensional entangled state over the two-dimensional entanglement in Sec.~II. We then derive the dissipative dynamics to prepare a three-dimensional entanglement in Sec.~III, and quantify the three-dimensional entanglement via negativity and fidelity in Sec.~IV. Subsequently we apply this dissipative mechanism to creation of a three-atom singlet state in Sec.~V.
This paper finally ends with a summary appearing in Sec.~VI.

\section{merits of multi-dimensional entanglement}
Before proceeding, we briefly review the advantage of the multi-dimensional entanglement. An arbitrary
$N$-dimensional entanglement can be written in the following form \cite{data}
\begin{equation}\label{f1}
|\Pi\rangle=\sum_{l=1}^Nc_l|l\rangle_A\otimes|l\rangle_B,
\end{equation}
where the subscripts $A$ and $B$ label two qudits, respectively and $|c_l|^2$ is the corresponding probability satisfying $\sum_{l=1}^N|c_l|^2=1$. This state is maximally entangled if $c_l=1/\sqrt{N}$. It has been proved in both theory and experiment that multi-dimensional entanglement can provide more security than two-dimensional entanglement offers. In the field of quantum key distribution for instance, the maximal
admissible error rate is shown
to increase by 53.48\% from the two-dimensional entanglement based protocol $E_A=1-F_A=1-(1/2+1/\sqrt{8})\simeq14.64\%$ to the three-dimensional entanglement based protocol 22.47\% \cite{chenjl1}, and this value further increases with the dimensionality $N$ \cite{chenjl}. In addition, the multi-dimensional entanglement $|\Pi\rangle$, acting as a quantum channel for high-dimensional quantum communication, quantum teleportation and other quantum information protocols, violates the local realism more strongly compared with the two-dimensional entanglement \cite{ly}. Therefore, the physical realization of multi-dimensional entanglement is at the request of quantum information processing \cite{mair,vaziri,vaziri1,Torre}.
\section{dissipative dynamics}
\begin{figure}
\centering\scalebox{0.23}{\includegraphics{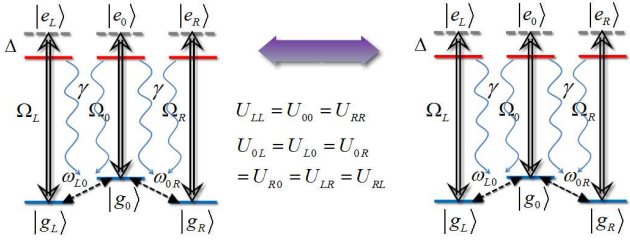} }
\caption{\label{p0}(Color online) Level diagram of two identical Rydberg atoms. The transition from ground state $|g_{L(0,R)}\rangle$
to the Rydberg state $|e_{L(0,R)}\rangle$ is driven by a classical field with Rabi frequency $\Omega_{L(0,R)}$, detuned by $-\Delta$. The resonant couplings between ground states are realized by two microwave fields with Rabi frequencies $\omega_{L0}$ and $\omega_{0R}$ respectively. For convenience, we assume the Rydberg interaction strengths  satisfy ${\cal U}_{LL}={\cal U}_{00}={\cal U}_{RR}=2{\cal U}$ and ${\cal U}_{0L}={\cal U}_{L0}={\cal U}_{0R}={\cal U}_{R0}={\cal U}_{LR}={\cal U}_{RL}={\cal U}$. The excited state can spontaneously decay into three ground states with branching ratio $\gamma/3$.}
\end{figure}
The system for dissipative preparation of the three-dimensional entanglement is illustrated in Fig.~\ref{p0}. We consider two Cesium atoms, where the ground levels $|g_L\rangle$, $|g_0\rangle$ and $|g_R\rangle$
correspond to hyperfine ground atomic levels $|F=3,M=-1\rangle$, $|F=4,M=0\rangle$  and $|F=3,M=1\rangle$ of $6S_{1/2}$ manifold, and the three Rydberg states $|e_L\rangle=|F=3,M=-1\rangle$, $|e_0\rangle=|F=4,M=0\rangle$ and $|e_R\rangle=|F=3,M=1\rangle$ of $125P_{1/2}$. The transition from the ground states to the highly excited states are driven by three $\pi$ polarized lasers with Rabi frequencies $\Omega_L$, $\Omega_0$, and $\Omega_R$, detuning $-\Delta$.
The Rydberg mediated interaction arises from
the long-range van der Waals interaction proportional to $C_6/R^6$ with $R$ being the distance between two Rydberg atoms and $C_6$ depending on the quantum numbers of Rydberg state \cite{saff,tonylee}.  Since the Rydberg interaction energy depends on angular degrees of freedom, the coupling
strength will be different when $|e_L\rangle$, $|e_0\rangle$ and $|e_R\rangle$  belong to different Zeeman sublevels \cite{cabe}.
We here assume the Rydberg interaction strengths  satisfy ${\cal U}_{LL}={\cal U}_{00}={\cal U}_{RR}=2{\cal U}$ and ${\cal U}_{0L}={\cal U}_{L0}={\cal U}_{0R}={\cal U}_{R0}={\cal U}_{LR}={\cal U}_{RL}={\cal U}$, where ${\cal U}_{ij}$ means the Rydberg interaction strength when atoms are in the state $|e_ie_j\rangle$.  In a rotating frame, the master equation describing the system's dynamics reads
 \begin{equation}\label{H1}
\frac{d{\hat{\rho}}}{dt}=-\frac{i}{\hbar}[\hat{\cal H}_I,\hat{\rho}]+\sum_{n=1,2}\hat{\cal L}_n\hat{\rho} \hat{\cal L}_n^{\dag}-\frac{1}{2}(\hat{\cal L}_n^{\dag}\hat{\cal L}_n\hat{\rho}+\hat{\rho} \hat{\cal L}_n^{\dag}\hat{\cal L}_n),
\end{equation}
where $
\hat{\cal H}_I=\hat{\cal H}_1\otimes \hat{\cal I}_2+ \hat{\cal I}_1\otimes\hat{\cal H}_2+\hat{\cal V}
$ is the Hamiltonian of system
and $\hat{\cal L}_n$ is the Lindblad operator describing the decay processes of $n$th atom by spontaneous emission of light. For simplicity,  the excited states are assumed to spontaneously decay into three ground states with the same branching ratio $\gamma/3$. To avoid the population of other ground states via spontaneous emission, we may introduce $6p_{1/2}$ manifold of Cesium atom and apply a series of $\pi$ polarized lasers to recycle other ground states as the method adopted in Ref~\cite{carr}.
Explicitly, the one-atom operators and the Rydberg interaction operators expressed in the
basis $\{|g_L\rangle, |g_0\rangle, |g_R\rangle, |e_L\rangle,  |e_0\rangle,  |e_R\rangle\}$ are
\begin{eqnarray}\label{xxx1}
\hat{\cal H}_n=\left[\begin{array}{c c c c c c}
0 & \omega_{L0} &0 & \Omega_L & 0 & 0 \\
\omega_{L0}^* & 0 &\omega_{0R} & 0 & \Omega_0 & 0 \\
0 & \omega_{0R}^* &0 & 0 & 0 & \Omega_R \\
\Omega_L^* & 0 &0 & -\Delta & 0 & 0 \\
0 & \Omega_0^* &0 & 0 & -\Delta  & 0 \\
0 & 0 &\Omega_R^* & 0 & 0 & -\Delta  \\
\end{array}
\right],
\end{eqnarray}
and
\begin{eqnarray}\label{111}
\hat{\cal V}=\sum_{i,j=L,0,R}{\cal U}_{ij}|e_i\rangle_1\langle e_i|\otimes|e_j\rangle_2\langle e_j|,
\end{eqnarray}
where $\omega_{L0}(\omega_{0R})$ is the Rabi frequency of microwave field that couples ground states resonantly, which plays the important role in our approach.

The idea of dissipative preparation of three-dimensional entanglement can be generalized as follows: In the absence of microwave fields, the ground state $|g_ig_j\rangle$ will be firstly driven to the single-excitation subspace $\{|e_ig_j\rangle, |g_ie_j\rangle\}$ with detuning $-\Delta$,
then further
 excited to the bi-excitation $|e_ie_j\rangle$ with detuning ${\cal U}_{ij}-2\Delta$. Different from the usual Rydberg interaction to block all bi-excitation states, here we set $\Delta={\cal U}/2$. This choice will result in a resonant coupling between the ground states and the bi-excitation Rydberg state when both atoms are initially in different states ($i\neq j$), but an off-resonant interaction detuned by an amount of $\Delta$ as both atoms are excited to the same state ($i=j$). Thus combining with microwave driven fields and spontaneous emission from the Rydberg levels, the system will be pumped into a stationary three-dimensional entangled state.  In the regime of large detuning $\Delta\gg\Omega$, we can safely disregard the single-excitation subspaces and the effective Hamiltonian is obtained through the second-order perturbation theory:
\begin{eqnarray}\label{222}
\hat{\cal H}^{'}_{I}&=&\hat{\cal H}_{\Omega}+\hat{\cal H}_{\omega},\\
\hat{\cal H}_{\Omega}&=&\sum_{i\neq j}\frac{2\Omega^2}{\Delta}\bigg(|g_ig_j\rangle\langle e_ie_j|+{\rm H.c.}+|e_ie_j\rangle\langle e_ie_j|\nonumber\\&&+|g_ig_j\rangle\langle g_ig_j|\bigg)+\frac{2\Omega^2}{\Delta}\bigg(|\Psi\rangle\langle\Psi|+|\Phi\rangle\langle\Phi|
\nonumber\\&&+|\Upsilon\rangle
\langle\Upsilon|\bigg),\\
\hat{\cal H}_{\omega}&=&\bigg\{\frac{\omega}{\sqrt{2}}\bigg[(|g_Lg_0\rangle+|g_0g_L\rangle)(\sqrt{3}\langle \Phi|+\langle \Upsilon|)\nonumber\\&&+(|g_Rg_0\rangle+|g_0g_R\rangle)(\sqrt{3}\langle \Phi|-\langle \Upsilon|)\bigg]\nonumber\\
&+&\omega\bigg[(|g_0g_L\rangle+|g_Rg_0\rangle)\langle g_Rg_L|\nonumber\\&+&(|g_Lg_0\rangle+|g_0g_R\rangle)\langle g_Lg_R|\bigg]+{\rm H.c.}\bigg\}.
\end{eqnarray}
In the above expressions, we have assumed the Rabi frequencies of laser fields have the same real value $\Omega$, the Rabi frequency of the microwave field is $\omega$, and introduced the following states
\begin{eqnarray}\label{111}
&&|\Psi\rangle=\frac{1}{\sqrt{3}}(|g_Lg_L\rangle-|g_0g_0\rangle+|g_Rg_R\rangle),\\
&&|\Phi\rangle=\frac{1}{\sqrt{6}}(|g_Lg_L\rangle+2|g_0g_0\rangle+|g_Rg_R\rangle),\\
&&|\Upsilon\rangle=\frac{1}{\sqrt{2}}(|g_Lg_L\rangle-|g_Rg_R\rangle),
\end{eqnarray}
where $|\Psi\rangle$ is the desired three-dimensional entangled state to be prepared, and this state is capable of being transformed into Eq.~(\ref{f1}) for $N=3$ by subsequent singlet qubit operation or modulating the Rabi frequencies of microwave fields such that $\omega_{L0}=-\omega_{0R}=\omega$. The stark-shift term and coupling strengths in Eq.~(6) are obtained as
\begin{equation}\label{ww}
  \left\{
   \begin{aligned}
   &\frac{\langle g_ig_j|\hat{\cal H}^{r}_{\Omega}|S_{e_i,g_j}\rangle\langle S_{e_i,g_j}|\hat{\cal H}^{r}_{\Omega}|g_ig_j\rangle}{\Delta}=\frac{2\Omega^2}{\Delta} \\
   &\frac{\langle e_ie_j|\hat{\cal H}^{r}_{\Omega}|S_{e_i,g_j}\rangle\langle S_{e_i,g_j}|\hat{\cal H}^{r}_{\Omega}|e_ie_j\rangle}{\Delta}=\frac{2\Omega^2}{\Delta} \\
   &\frac{\langle g_ig_j|\hat{\cal H}^{r}_{\Omega}|S_{e_i,g_j}\rangle\langle S_{e_i,g_j}|\hat{\cal H}^{r}_{\Omega}|e_ie_j\rangle}{\Delta}=\frac{2\Omega^2}{\Delta} \\
   \end{aligned}
   \right.,
  \end{equation}
where $|S_{e_i,g_j}\rangle=(|e_ig_j\rangle+|g_ie_j\rangle)/\sqrt{2}$, and $\hat{\cal H}^{r}_{\Omega}$ is the Hamiltonian in the original interaction picture.

The corresponding Lindblad operators thus take the forms
\begin{eqnarray}\label{x1}
\hat{\cal L}_1^{L}&=&\sqrt{\frac{\gamma}{3}}\bigg[\sum_{i\neq j}(|g_L e_j\rangle
\langle e_i e_j|+
|g_L g_j\rangle\langle e_i g_j|
)\nonumber\\
&&+\bigg(\frac{1}{\sqrt{3}}|\Psi\rangle
+\frac{1}{\sqrt{6}}|\Phi\rangle
+\frac{1}{\sqrt{2}}|\Upsilon\rangle\bigg)\langle e_L g_L|\bigg],
\end{eqnarray}

\begin{eqnarray}\label{x2}
\hat{\cal L}_1^{R}&=&\sqrt{\frac{\gamma}{3}}\bigg[\sum_{i\neq j}(|g_Re_j\rangle
\langle e_i e_j|+
|g_R g_j\rangle\langle e_i g_j|
)\nonumber\\
&&+\bigg(\frac{1}{\sqrt{3}}|\Psi\rangle
+\frac{1}{\sqrt{6}}|\Phi\rangle
-\frac{1}{\sqrt{2}}|\Upsilon\rangle\bigg)\langle e_R g_R|\bigg],
\end{eqnarray}

\begin{eqnarray}\label{x3}
\hat{\cal L}_1^0&=&\sqrt{\frac{\gamma}{3}}\bigg[\sum_{i\neq j}(|g_0e_j\rangle\langle e_ie_j|+
|g_0g_j\rangle\langle e_ig_j|)\nonumber\\
&&+\bigg(\frac{2}{\sqrt{6}}|\Phi\rangle
-\frac{1}{\sqrt{3}}|\Psi\rangle\bigg)\langle e_0g_0|\bigg],
\end{eqnarray}

\begin{eqnarray}\label{x4}
\hat{\cal L}_2^{L}&=&\sqrt{\frac{\gamma}{3}}\bigg[\sum_{i\neq j}(|e_jg_L \rangle
\langle e_je_i |+
|g_jg_L \rangle\langle g_je_i |
)\nonumber\\
&&+\bigg(\frac{1}{\sqrt{3}}|\Psi\rangle
+\frac{1}{\sqrt{6}}|\Phi\rangle
+\frac{1}{\sqrt{2}}|\Upsilon\rangle\bigg)\langle g_Le_L |\bigg],
\end{eqnarray}

\begin{eqnarray}\label{x5}
\hat{\cal L}_2^{R}&=&\sqrt{\frac{\gamma}{3}}\bigg[\sum_{i\neq j}(|e_jg_R\rangle
\langle  e_je_i|+
| g_jg_R\rangle\langle  g_je_i|
)\nonumber\\
&&+\bigg(\frac{1}{\sqrt{3}}|\Psi\rangle
+\frac{1}{\sqrt{6}}|\Phi\rangle
-\frac{1}{\sqrt{2}}|\Upsilon\rangle\bigg)\langle g_Re_R |\bigg],
\end{eqnarray}

\begin{eqnarray}\label{x6}
\hat{\cal L}_2^0&=&\sqrt{\frac{\gamma}{3}}\bigg[\sum_{i\neq j}(|e_jg_0\rangle\langle e_je_i|+
|g_jg_0\rangle\langle g_je_i|)\nonumber\\
&&+\bigg(\frac{2}{\sqrt{6}}|\Phi\rangle
-\frac{1}{\sqrt{3}}|\Psi\rangle\bigg)\langle g_0e_0|\bigg],
\end{eqnarray}
where $\hat{\cal L}_n^{j}$ denotes the spontaneous decay of $n$th atom from the excited states to the ground state $|g_j\rangle$.

\section{measure of three-dimensional entanglement}
\subsection{Negativity}
A steady-state solution $\hat{\rho}_{\infty}$ of system should satisfy the following equation
\begin{equation}\label{H9}
0=-\frac{i}{\hbar}[\hat{\cal H}^{'}_I,\hat{\rho}_{\infty}]+\sum_{n=1,2}\hat{\cal L}_n\hat{\rho}_{\infty} \hat{\cal L}_n^{\dag}-\frac{1}{2}(\hat{\cal L}_n^{\dag}\hat{\cal L}_n\hat{\rho}_{\infty}+\hat{\rho}_{\infty} \hat{\cal L}_n^{\dag}\hat{\cal L}_n).
\end{equation}
The evolution of system is totally governed by the coaction of unitary dynamics of Eq.~(\ref{222}) and dissipative dynamics from Eq.~(\ref{x1}) to Eq.~(\ref{x6}). It is easy to see that if the atoms are initially prepared in different ground states, they will be simultaneously pumped into the bi-excitation state and then decay into one of the ground state subspaces with both atoms in the same state $\{|\Psi\rangle,|\Phi\rangle,|\Upsilon\rangle\}$. The microwave fields of Hamiltonian $\hat{\cal H}_{\omega}$ then drive $|\Phi\rangle$ and $|\Upsilon\rangle$ into $|g_Lg_0\rangle(|g_0g_L\rangle)$ or $|g_Rg_0\rangle(|g_0g_R\rangle)$ and the process of pumping and decaying repeats again, while leaves the three-dimensional entangled state $|\Psi\rangle$ invariant. Therefore the final population from an arbitrary initial state will be accumulated in the state $|\Psi\rangle$, i.e. the three-dimensional entangled state $\hat{\rho}_{\infty}=|\Psi\rangle\langle \Psi|$ is the unique solution of Eq.~(\ref{H9}).
\begin{figure}
\scalebox{0.42}{\includegraphics{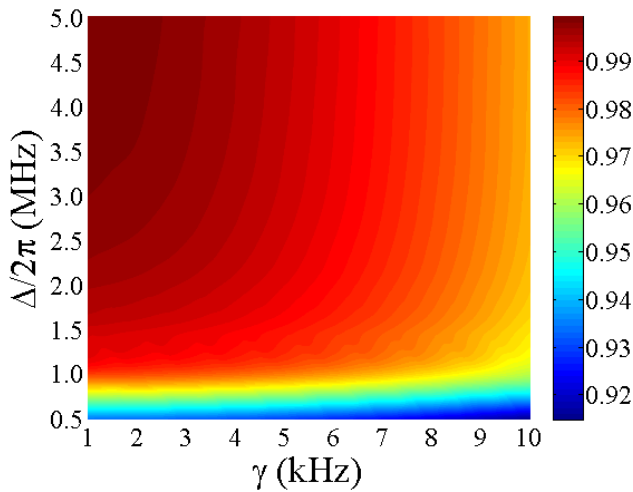} }
\caption{\label{pj}(Color online) The negativity ${\cal N}(\hat\rho_{\infty})$ versus  single-photon detuning parameter $\Delta$ and spontaneously decay rate $\gamma$. Other parameters: $\Omega/2\pi$=0.02~MHz, ${\cal U}=2\Delta$, $\omega=3\Omega^2/(4\Delta)$.}
\end{figure}
To effectively measure this kind of three-dimensional entanglement of two particles,  we need to introduce the trace norm of the partial transposition $||\hat\rho_{\infty}^{T_1}||={\rm Tr}\sqrt{\hat\rho_{\infty}^{T_1\dag}\hat\rho_{\infty}^{T_1}}$, where $\hat\rho_{\infty}^{T_1}$ denotes the
partial transposition of $\hat\rho_{\infty}$ with respective to the subsystem of the first atom with basis adopted in Eq.~(\ref{xxx1}).
The Peres-Horodecki criterion of separability shows that if $\hat\rho_{\infty}^{T_1}$ is not positive, then  $\hat\rho_{\infty}$ is not separable. The negativity is defined as \cite{vidal}
\begin{equation}\label{ss}
{\cal N}(\hat\rho_{\infty})\equiv\frac{||\hat\rho_{\infty}^{T_1}||-1}{2},
\end{equation}
which is equal to the absolute value of the sum of negative
eigenvalues of $\hat\rho_{\infty}^{T_1}$. In the ideal case, the negativity of a pure three-dimensional entangled state takes the value of unity. In Fig.~\ref{pj}, we characterizes the dynamical evolution of negativity ${\cal N}(\hat\rho_{\infty})$ versus the single-photon detuning parameter $\Delta$ and the spontaneously decay rate $\gamma$ by numerically solving the steady-state equation (\ref{H9}).
It is shown that the negativity keeps high over a wide range of parameters. It approaches to unity as the increase of $\Delta$, and the maximum three-dimensional entanglement is found to be 99.91\% at $\Delta/2\pi=5.0$~MHz, $\gamma=1$~kHz within the given range of parameters. This result is in accordance with the approximation made in Eq.~(\ref{222}), since a larger value of $\Delta$ is capable of guaranteeing the quantum state with two atoms in the same ground state more stable during the pumping process.

\subsection{Fidelity}

\begin{figure}
\scalebox{0.33}{\includegraphics{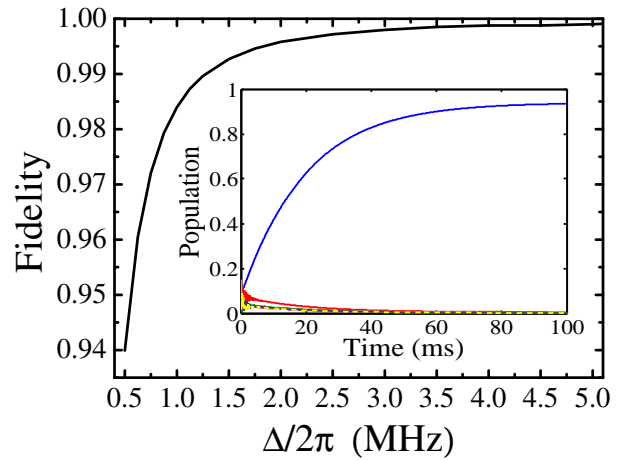} }
\caption{\label{p1}(Color online) The dependence of steady-state fidelity ${\cal F}(\hat\rho_{\infty})=\langle \Psi|\hat{\rho}_{\infty}|\Psi\rangle$ on the single-photon detuning parameter $\Delta$. Other parameters: $\Omega/2\pi$=0.02~MHz,$\gamma$=1~kHz, ${\cal U}=2\Delta$, $\omega=3\Omega^2/(4\Delta)$. The inset demonstrates the time evolutions of populations for an initially mixed state corresponding to $\Delta/2\pi$=0.5~MHz.}
\end{figure}
 Another simple and efficient way to assess the quality of entangled state is the fidelity.  In Fig.~\ref{p1}, we plot the dependence of steady-state fidelity ${\cal F}(\hat\rho_{\infty})=\langle \Psi|\hat{\rho}_{\infty}|\Psi\rangle$ on the single-photon detuning parameter $\Delta$, provided other parameters are given by $\Omega/2\pi$=0.02~MHz,$\gamma$=1~kHz,  $\omega=3\Omega^2/(4\Delta)$.
 Corresponding to the negativity, the fidelity exhibits asymptotic behaviour as the increase of $\Delta$.
The inset of Fig.~\ref{p1} further demonstrates the time evolutions of populations for an initially mixed state with $\Delta/2\pi$=0.5~MHz, which signifies a fidelity exceeding 90\% is achievable in a short interaction time 100~ms. It is worth noting that the proposed mechanism is much more like the cooling mechanism in Refs.~\cite{vacanti,busch}, where the laser sideband cooling is applied to induce a relatively large detuning of the desired
state but resonant lasers drive all other qubit states. Therefore, our method can also be reconsidered as a cooling process that the whole system is finally cooled into the three-dimensional entangled state at the coaction of Rydberd interaction and spontaneous emission.

\section{generalization to tripartite singlet state}
\begin{figure}
\scalebox{0.5}{\includegraphics{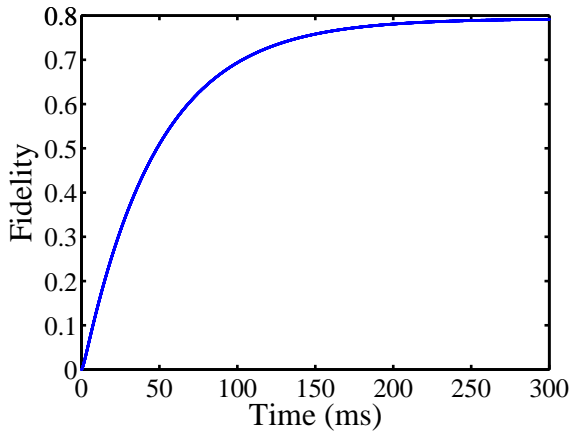} }
\caption{\label{pz}(Color online) The fourth-order Runge-Kutta numerical  simulation of fidelity for creation of the three-atom singlet state $|S_3\rangle$, corresponding to the parameters $\Omega/2\pi$=0.02~MHz,$\gamma$=1~kHz, ${\cal U}=0.2\Delta$, $\omega=3\Omega^2/(4\Delta)$, and $\Delta/2\pi$=0.5~MHz. The initial state is selected as $|g_Lg_Lg_L\rangle$ and  the fidelity approaches around 80\% after 300 ms. }
\end{figure}
The considered configuration of atoms not only is successful in the above bipartite entanglement, but also may be helpful for multipartite entanglement. An interesting  type of entangled state called $N$-particle
$N$-level singlet states, can be expressed as \cite{cab}
\begin{equation}\label{H}
|S_N\rangle=\frac{1}{\sqrt{N!}}\sum_{\{n_l\}}\epsilon_{n_1,\cdots,n_N}|\alpha_{n_1},\cdots,\alpha_{n_N}\rangle,
\end{equation}
where $\epsilon_{n_1,\cdots,n_N}$ is the generalized Levi-Civita
symbol, and the state $|\alpha_{n_{i}}\rangle$ denotes one base of
the qudit. These entanglement states related to violations of Bell's inequalities, are the
key resource for the solutions of ``$N$-strangers", ``secret
sharing" and ``liar detection" problems, which have no classical
solutions. Now we exploit the possibility to achieve the three-atoms singlet state  through the  same dissipative-pumping mechanism.
To be concrete, the form of a three-atoms singlet state in connection with the current model reads
\begin{eqnarray}\label{111}
|S_3\rangle&=&\frac{1}{\sqrt{6}}(|g_0g_Lg_R\rangle-|g_Lg_0g_R\rangle-|g_Rg_Lg_0\rangle
+|g_Lg_Rg_0\rangle\nonumber\\&&+|g_Rg_0g_L\rangle-|g_0g_Rg_L\rangle).
\end{eqnarray}
To complete this task, we may set the Rydberg interaction ${\cal U}_{LL}={\cal U}_{00}={\cal U}_{RR}=2\Delta$ and ${\cal U}_{0L}={\cal U}_{L0}={\cal U}_{0R}={\cal U}_{R0}={\cal U}_{LR}={\cal U}_{RL}={\cal U}\ll \Delta$. This arrangement guarantees a resonant transition between two atoms when they are in the same state, but a detuned interaction as their states are different. Therefore the atoms will be stabilized in the ground states that each pair of atoms populate in different ground states. Most importantly, the microwave fields leave $|S_3\rangle$ unchanged but are coupled to other states transition. Finally, the population of three-atoms singlet state will be accumulated into a high value. As shown in Fig.~\ref{pz}, the evolution of fidelity for this three-atom singlet state versus time is plotted by  numerical simulation via fourth-order Runge-Kutta, where the target state fidelity is stabilized at 79.15\% after 300 ms. This value
may be low for quantum computation but are relatively high in
the sense of steady-state entanglement.

\section{summary}
Nowadays, the Rydberg atom has been exploited to many fields of quantum information due to the availability of a strong, long-range interaction between Rydberg atoms with principal quantum number $n\gg$ 1. In this work, we have put forward an efficient way for generation of stationary three-dimensional entangled state by combing the coherent dynamics  provided by Rydberg mediated interaction and the dissipative dynamics induced by spontaneous emission of Rydberg state. Both the negativity and fidelity are nearly perfect under the given parameters of Cesium atom. This dissipative mechanism can be also generalized to creation of a three-atom singlet state.
We believe that our work may open a new avenue for the entanglement preparation experimentally in the near future.

\begin{center}{\bf{ACKNOWLEDGMENT}}
\end{center}
The authors thank the anonymous reviewer for constructive
comments that helped in improving the quality of this paper.
This work is supported by Fundamental Research Funds for the Central Universities under Grant No. 12SSXM001,  National
Natural Science Foundation of China under Grant Nos. 11204028 and 11175044, and National Research
Foundation and Ministry of Education, Singapore (Grant
No. WBS: R-710-000-008-271). X. Q. Shao is also supported in part by the Government of China through CSC.
\

\end{document}